\documentclass[twocolumn,showpacs,amsmath,amssymb,prl,a4paper]{revtex4}
\usepackage{graphicx}
\usepackage{dcolumn}
\usepackage{bm}
\newcommand{\PP}{\mathcal{P}}

\newcommand{\Z}{\mathcal{Z}}
\newcommand{\F}{\mathcal{F}}

\newcommand{\psii}{\psi_{i}}
\newcommand{\psif}{\psi_{f}}

\newcommand{\sech}{\mathop\mathrm{sech} }
\newcommand{\Sim}{\mathop{\sim}}

\newcommand{\rmi}{i}
\newcommand{\rme}{e}
\newcommand{\Or}{\mathrm{O}}

\renewcommand{\case}{\frac}

\begin{document}

\title{Exact solution of the nonlinear laser passive mode locking transition}
\author{Omri Gat, Ariel Gordon and Baruch Fischer}
\address{Department of Electrical Engineering, Technion, Haifa
32000, Israel }

\date{\today}
\begin{abstract}
We present the first statistical mechanics study of a passively
mode locked laser which includes all the main physical processes,
saturable absorption, Kerr nonlinearity, parabolic gain filtering
and group velocity dispersion, assuming the soliton condition. We
achieve an exact solution in the thermodynamic limit, where the
ratio of the cavity length to the pulse width, the duty cycle,
tends to infinity. The thermodynamics depends on a single
dimensionless parameter $\gamma$, the ratio of the correlation
length to the pulse width. The phase diagram consists of one
ordered, mode-locked phase and one disordered, continuous wave
phase, separated by a first order phase transition at $\gamma=9$.
The model belongs to a new class of solvable statistical
mechanics models with a non-trivial phase diagram. The results
are obtained with a fully controlled transfer matrix
calculation, showing rigorously that passive mode locking is a
thermodynamic phase transition.
\end{abstract}

 \maketitle



\paragraph{Introduction}
Formation of ultrashort pulses in lasers by passive mode locking
is an important branch of optics, both
from the point of view of basic research and of practical
applications. As such, it has been the subject of many theoretical
and experimental studies over several decades, see
\cite{haus-review} for a review. Nevertheless, until recently the
central question of the threshold power needed to trigger passive
mode locking was standing without a satisfactory answer.

In several recent works \cite{GFPRL,GGF03,GFOC} considerable
progress was made toward resolving this issue by including the
effect of \emph{noise} in a nonperturbative manner. The addition
of a random element into the cavity electrodynamics turns the
laser into a statistical physics system. Applying methods of
equilibrium statistical mechanics, it has been shown that
sufficiently strong noise destabilizes the pulses formed by a
saturable absorber, and that the process of pulse formation is a
first order phase transition.

However, these results were all obtained in the context of models
of the laser dynamics which assumed a simple form of spectral gain
filtering. Here, for
the first time, we tackle the full problem of passively
mode-locked laser with a fast saturable absorber, quadratic gain
filtering, slow saturable gain, group velocity dispersion, and
Kerr nonlinearity, assuming the soliton condition.

In the first step of the theoretical analysis we construct an
exact mapping from the statistical steady state of the laser
electric field to the thermal fluctuations of a string with an
unstable self-interaction. We show that the
thermodynamic limit is obtained when the laser cavity is much
longer than one of the natural length scales, the pulse width or
the correlation length. It is next shown that thermodynamics
is determined solely by the dimensionless ratio $\gamma$ of the
two natural length scales. We proceed to calculate exactly and
explicitly the free energy, first by physical, mean field-like
arguments, which are then established by a rigorous transfer
matrix calculation. The free energy enables us
to calculate all thermodynamic quantities, including pulse power
as a function of $\gamma$, and the thermodynamic phase diagram,
which consists of one ordered, mode locked phase for
$\gamma>9$, and one disordered, non-mode locked phase for
$\gamma<9$. Pulse formation is a first order phase transition. The pulsed configuration is metastable for $8<\gamma<9$, while the continuous wave configuration is metastable for all $\gamma>9$.

The model closely resembles equations which have been used extensively to study kinetics of phase transitions and critical phenomena \cite{hh}. However, the opposite sign of the nonlinearity leads to markedly different phenomenology. The passive mode locking equation therefore belongs to a new class of statistical physics models.

\paragraph{Thermodynamics~of~passively~mode~locked~lasers.}
\hspace{-2mm}Our starting point is the master equation which governs the slow
dynamics of the complex envelope of the electric field $\psi(x,t)$
in a cavity of length $L$ \cite{haus-review,GFPRL}
\begin{equation}\label{eq:motion}
\partial_t\psi=(\gamma_g+\rmi\gamma_d)\partial_z^2\psi
+(\gamma_s+\rmi\gamma_k)|\psi|^2\psi+g\psi+\eta\ .
\end{equation}
in $0\le z\le L$ with periodic boundary condition, where the real
consants $\gamma_g>0$, $\gamma_d$, $\gamma_s>0$, $\gamma_k$ are
the coefficients of spectral filtering, group velocity dispersion,
(fast) saturable absorption, and Kerr nonlinearity, respectively.
Noise of spontaneous emission and other sources is modelled by
the random term $\eta$, which is a (complex) Gaussian process
with covariance
$\left<\eta^*(x,t)\eta(x',t')\right>=2TL\delta(x-x')\delta(t-t')$.
Finally, as shown in \cite{GGF03}, the slow saturable gain $g$,
may be chosen, without significant loss of generality, such that
it sets the total intracavity power
$\|\psi\|^2=\frac1L\int_0^Ldx|\psi(x)|^2$ to a fixed value $P$.
$g$ becomes then a Lagrange multiplier for the fixed power constraint.

\begin{figure}[htb]
\hbox{\hspace{-.2cm}\includegraphics[width=8.8cm]{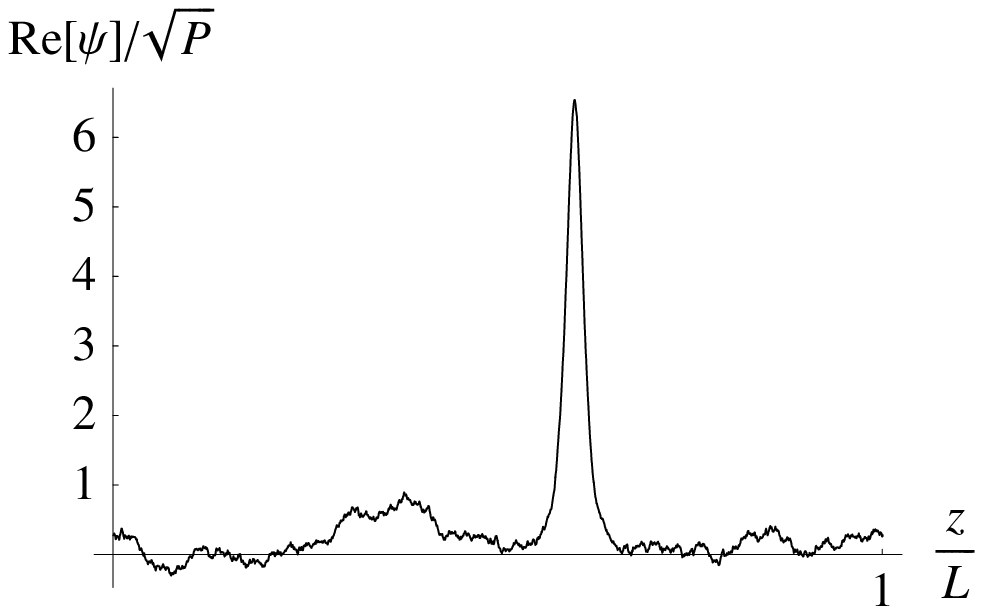}}\vspace{2mm}
\hbox{\hspace{-1.cm}\includegraphics[width=10.5cm]{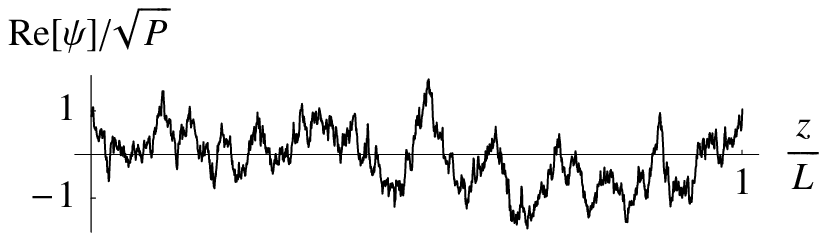}}
\caption{A typical realization of the real part of the envelope
$\psi$ of the electric field in the laser cavity, in the mode
locked phase (top) and in the continuous wave phase
(bottom).\label{fig:pulse}}
\end{figure}

In this Letter we consider Eq.\ (\ref{eq:motion})  in the special
but important case that ${\gamma_s+\rmi\gamma_k}$ is a real
multiple of ${\gamma_g+\rmi\gamma_d}$, known as the
{soliton} condition. In this case one can define a `Hamiltonian'
functional \cite{GFOC}
\begin{equation}\label{eq:h}
H[\psi]=\int_0^L dx\,
(-\case12\gamma_s|\psi(x)|^4+\gamma_g|\psi'(x)|^2)\ ,
\end{equation}
such that the invariant measure $\rho[\psi]$ of equation
(\ref{eq:motion}) is \hspace{-5mm}\vbox{
\begin{eqnarray}\label{eq:invariant}
&&\rho[\psi]=Z^{-1}\rme^{-H[\psi]/(LT)}\delta(P-\|\psi\|^2)\ ,\\
\label{eq:z} &&Z=\int[d\psi][d\psi^*]
\rme^{-H[\psi]/(LT)}\delta(P-\|\psi\|^2)\ .
\end{eqnarray}}
Note that the power constraint is enforced explicitly, and that
$\rho$ is independent of the imaginary terms in Eq.\ 
(\ref{eq:motion}). The study of steady state preoperties of
Eq.\ (\ref{eq:motion}) is now reduced, as in
\cite{GFPRL,GGF03}, to the that of an equilibrium statistical
mechanics system with partition function $Z$.

The Hamiltonian functional $H$ is almost identical to the critical Ginzburg-Landau (GL) functional, the paradigm for the effective description of continuous phase transition \cite{hh}. In Eq.\ (\ref{eq:h}), however, in contrast with the GL functional, the coefficient of the quartic term is \emph{negative}, making the null configuration unstable. The instability is countered by the power constraint, which acts nonlocally, and the thermodynamics is consequently radically different from the GL one. In particular, the one-dimensional system exhibits an ordering transition, which is impossible in the one-dimensional GL model.

The character of the steady state distribution
$\rho$ is determined by the strength of the
ordering saturable absorber relative to the disordering noise.
The system has two natural length scales: The \emph{pulse width}
$L_p=\frac{4\gamma_g}{\gamma_sPL}$ measures the effect of
saturable absorption, while the \emph{correlation length}
$L_c=\frac{\gamma_gP}{LT}$ measures the effect of noise. In both
cases, the effect is stronger the smaller is the associated
length scale. Thus, $L_p$ is smaller the larger is $\gamma_s$ and
is independent of $T$, while the converse is true for $L_c$. The
third length scale in the system, the cavity length $L$, is
typically much larger than $L_p$ and $L_c$ in multimode lasers;
we therefore study the statistical problem in the limit $L\gg
L_p,L_c$, which serves as the thermodynamic limit,
 neglecting small corrections of $\Or(L_p/L)$ or $\Or(L_c/L)$.
Thermodynamic quantities, and in particular the mode locking
threshold are therefore determined by the sole dimensionless
parameter, the ratio $\gamma=4L_c/L_p=\gamma_sP^2/T$. Note that
thermodynamics is independent of the spectral filtering
$\gamma_g$.

Our analysis proceeds in the textbook approach of calculating the
free energy $F=-\log Z$ of the statistical mechanics problem
Eqs.~(\ref{eq:invariant}-\ref{eq:z}), from which other
thermodynamic quantities follow. However, seeking to calculate
the partition function we run into a common obstacle, namely that
the functional integral in Eq.\ (\ref{eq:z}) is not well defined. Mathematically this is not a serious problem, since the
invariant measure \emph{is} well-defined \cite{stratonovich}, but
in order to use $Z$ and $F$ we need to give a precise meaning to
the functional integral, as the continuum limit of a regularized
version where the integration is finite-dimensional. Given a regularization scheme with $N$ $\psi$ integration, we define the regularized partition function
\begin{equation}\label{eq:zn}
Z_N=\int\prod_{n=1}^N
({\textstyle\frac{a_N\gamma_g}{L^2T}}d\psi_nd\psi^*_n)
\rme^{-\frac{H_N[\psi]}{LT}}\delta(P-\|\psi\|_N^2)
\end{equation}
where $H_N$ and $\|\cdot\|_N$ are regularized versions of $H$ and
$\|\cdot\|$. The integration measure is multiplied by factors of
$\frac{\gamma_g}{LT}$ to make $Z_N$ dimensionless, and by factors
of $a_N$, a regularization scheme- and $N$-dependent
dimensionless constant, to make $Z=\lim_{N\to\infty}Z_N$ finite. The limit is independent of the
regularization scheme up to an unimportant multiplicative
constant.

The rest of this Letter is devoted to the calculation of $F$, Eq.\ (\ref{eq:f}), and from it the phase diagram and
thermodynamic quantities, which all have simple algebraic expressions. It is shown that when $\gamma>9$ the equilibrium is an ordered, mode-locked phase where the power $P$ is divided between a single pulse
and continuum fluctuations, see the top panel of Fig.\ \ref{fig:pulse}; when $\gamma<9$ the equilibrium is a disrodered phase for, where the electric field consists only of spatially
homogeneous fluctuations, see the bottom panel of
Fig.~\ref{fig:pulse}.

\paragraph{Mean field calculation of the free enegy}
As a preliminary step towards the calculation of $F$ we examine
the problem in two solvable limits. In the first, $T\to0$, the
evolution equation (\ref{eq:motion}) approaches the
noiseless dynamics without the random term $\eta$. Then, as is well
known, $\psi$ reaches an equilibrium in the form of a soliton-like pulse
$\psi_p(z)=\rme^{\rmi\phi}\sqrt{\frac{PL}{2L_p}}
\sech\Big(\frac{z-z_0}{L_p}\Big)$, with two initial
conditions-dependent real parameters, the pulse position $0\le
z_0<L$ and phase $0\le\phi<2\pi$.

In our context, $T\to0$ is the validity condition for the Laplace
method, otherwise known as the saddle point approximation, in Eq
(\ref{eq:z}), and the solitons $\psi_p$ take the role of the
saddle points.
It follows that $Z\Sim\limits_{T\to0}\rme^{-H[\psi_p]/(LT)}$, so
that the free energy when $L_c\gg L_p$ or $\gamma\gg1$ is
\begin{equation}\label{eq:Fp}
F_{p}=\frac{H[\psi_p]}{LT}=-\frac{\gamma_s^2L^2P^3}{48\gamma_gT}
=-\frac{LL_c}{4L_p^2}\ .
\end{equation}

In the second solvable limit $\gamma_s\to0$, $H$ becomes
quadratic and, after replacing the power constraint delta
function by its Fourier integral representation, the functional
integral in Eq.\ (\ref{eq:z}) becomes gaussian. The quadratic form
is then diagonalized using the Fourier representations of $\psi$,
and we calculate $Z$ in a regularization scheme where $N$ is the
number of $\psi$ Fourier modes kept (see Eq.\ (\ref{eq:zn})). The
result of the gaussian integration is
\begin{equation}\label{eq:zcr}
Z_{c}=
\int_{-\rmi\infty}^{\rmi\infty}\frac{dw}{2\pi\rmi}\,\rme^{z}
\!\prod_{n=-\infty}^\infty \bigg(1+\frac{Lw}{L_c(2\pi
n)^2}\bigg)^{-1}\ .
\end{equation}
Under the assumption that $L\gg L_c$, we can use the standard
approximation methods of Euler-MacLaurin summation and saddle
point integration to get
the free energy in the limit $L_c\ll L_p$ or $\gamma\ll1$
\begin{equation}\label{eq:fc}
F_{c}=\frac{L}{4L_c}=\frac{L^2T}{4\gamma_gP}\ .
\end{equation}

Although the preceding expressions Eqs (\ref{eq:Fp},\ref{eq:fc}) for the free energy were obtained in limiting
cases, we now argue that $F_p$, the pulse free energy, and $F_c$,
the continuum free energy, may combined  into an expression for $F$ valid
for every $\gamma$. The argument is based on the following
assumption: Configurations $\psi$ which contribute significantly
to $Z$ are such that $\psi(z)=\Or(1)$ for most $z$, with possibly
few narrow regions where $\psi(z)=\Or(\sqrt{L/L_p})$, whose total
width is $\Or(L_p)$. Let $yP$, $0\le y\le1$ be the total power
concentrated in regions where $\psi$ is large. For $z$ values
where $\psi$ is small the nonlinear term in $H$ is negligible,
and the existence of regions of large $\psi$ affects the
statistics of the small $\psi$ region only in that the total
available power for fluctuations is $(1-y)P$ rather than $P$.
The regions of small $\psi$ therefore contribute
$F_c|_{P\to(1-y)P}$ to the total free energy. Similarly, the
regions of large $\psi$ are so narrow that noise induced
fluctuations make negligible contribution to the free energy in
them, so that the large $\psi$ regions
contribute $H[\psi]/(LT)$ to $F$. By
the principle that $F$ is minimized by the Gibbs distribution the large $\psi$ regions should be such that $H[\psi]$
is minimized, that is, $\psi$ will assume a soliton-like shape
with total power $yP$, and contribute $F_p|_{P\to yP}$ to $F$. We
claim that since the small parameter in these arguments is
$L_p/L$, they become exact in thermodynamic limit.

The total free energy is the sum of the pulse and continuum
parts, minimized over values of $y$,
\begin{eqnarray}
F=\min_y\bigg( -\frac{\gamma_s^2L^2(yP)^3}{48\gamma_gT}+
\frac{L^2T}{4\gamma_gP(1-y)}\bigg)\nonumber\\\label{eq:f}
=\min_y\frac{L}{L_p}\bigg(-\frac{\gamma y^3}{12}+
\frac{1}{\gamma(1-y)}\bigg)\ .
\end{eqnarray}
$F$ can be used to easily derive other thermodynamic quantities.
An order parameter which is nonzero if and only
if mode locking occurs is $M\equiv \frac{L_p}{L}
\sqrt{\left<{|\psi|^4}\right>}$. $M$ has dimensions of power, and
is proportional to the experimentally measurable RF power
\cite{experiment}. It follows from Eq.~(\ref{eq:zn}) and the
definition of $F$ that
$\left<|\psi|^4\right>=-2T\partial_{\gamma_s}F$. Letting
$\Phi(\gamma,y)$ denote the target function in Eq.\ 
(\ref{eq:f}), and $\bar y(\gamma)$ the minimizer, we calculate
\begin{equation}\label{eq:mgamma}
M(\gamma)= -2T(L_p/L)
\sqrt{\partial_{\gamma_s}\Phi(\gamma,\bar y(\gamma))}
=({{\bar y^3}/3})^{\frac12}P\ .
\end{equation}
\begin{figure}[tb]
\includegraphics[width=8cm]{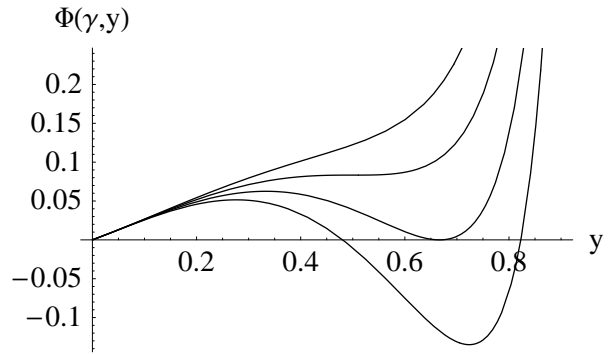}
\caption{The function $\Phi(\gamma,y)$ whose minimum with respect
to $y$ is the free energy as a function $y$ for
$\gamma=7,8,9,10$ with higher values of $\gamma$ corresponding to
lower curves. Curves with two minima correspond to systems with a
metastable state. When $\gamma=9$ the
values of $\Phi(\gamma,\cdot)$ at the two minima are
equal.\label{fig:fgamma}}
\end{figure}
The most important consequence of equation (\ref{eq:mgamma}) is
that mode locking occurs whenever the minimizer 
$\bar y(\gamma)$ is greater than zero. $\Phi(\gamma,\cdot)$
has the following behavior: For $\gamma\le8$ it has a single
minimum at $0$, while for $\gamma>8$ there is a second minimum
at $\frac12\big(1+\sqrt{1-8/\gamma}\big)$, which
becomes a global minimum when $\gamma\ge9$, see Fig.~\ref{fig:fgamma}. Therefore, as
$\gamma$ is increased through $\gamma=9$, $M$ jumps
discontinuously from $0$ to $\frac{\sqrt{8}}{9}P$; that is, the
transition is first order, the ordered phase is metastable when
$8<\gamma<9$, and the disordered phase is metastable for all $\gamma>9$. The pulse power in the mode locked phase is
$\bar y(\gamma)P$, as can be verified by calculating higher moment.
\nopagebreak
\paragraph{Transfer matrix calculation of the free energy}
We turn to the transfer matrix calculation of $Z$ which
establishes our expression (\ref{eq:f}) for the free energy with
a controlled derivation. For this purpose we consider a slightly
generalized partition function $\Z$ with fixed
values of $\psi$ at the endpoints of the interval,
$\psi(0)=\psii$, $\psi(L)=\psif$, and generalized free energy
$\F=-\log\Z$. In this calculation it is more convenient to use
system parameters that do not depend on the system size, so we
define
\begin{widetext}
\begin{equation}\label{eq:zif}
\Z(\psii,\psif,\alpha,\beta,\PP,L)
=\int_{\psi(0)=\psii}^{\psi(L)=\psif}[d\psi][d\psi^*]
\times\rme^{\int_0^Ldx\,(\frac12\alpha|\psi(x)|^4-\beta|\psi'(x)|^2)}
\delta\Big(\int_0^Ldx\,|\psi(x)|^2-\PP\Big)\ .
\end{equation}
Like $Z$, $\Z$ needs to be defined using a limiting procedure
with a properly scaled functional measure. Let $\Z_\delta$ denote
the lattice regularization of $\Z$ with lattice spacing $\delta$.
It satisfies the identity
\begin{equation}\label{eq:ztr}
\Z_\delta(\psii,\psif,\alpha,\beta,\PP,L)= \int
\!\frac{\beta}{\delta\pi}d\psi d\psi^*
\Z_\delta(\psii,\psi,\alpha,\beta,\PP-\delta|\psi|^2,L-\delta)
\rme^{\frac{\delta}{2}\alpha|\psi|^4-\frac{\beta}{\delta}|\psif-\psi|^2}
\end{equation}
\end{widetext}
We next expand the right-hand-side in powers of $\delta$ and
$\psi-\psif$ and perform the gaussian integration. This gives an
equation involving $\Z_\delta$ and its derivatives. Taking then
the continuum limit $\delta\to0$ gives the equation for $\Z$
\begin{equation}\label{eq:zdiff}
\case1\beta\partial_{\psif}\partial_{\psif^*}\Z-
|\psif|^2\partial_\PP\Z-\partial_L\Z+\case{1}2\alpha|\psif|^4\Z=0
\ ,
\end{equation}
and from it the equation for $\F$
\begin{equation}\label{eq:fdiff}
\frac1\beta|\partial_{\psif}\F|^2-|\psif|^2\partial_\PP\F+\case{1}2\alpha|\psif|^4=0\ ,
\end{equation}
where the term with double $\psif$ derivative and the term with
an $L$ derivative were dropped, since they are $\Or(L_p/L)$ with
respect to the terms retained.

Mean field arguments, similar to those presented above to obtain
$F$, can be used to calculate $\F$. Like $F$, $\F$ is the sum of
a continuum contribution $\F_c$ where the nonlinear term is
unimportant, and a pulse term $\F_p$ where noise is unimportant.
Since the continuum fluctuations are a bulk property, $\F_c$ is
independent of the boundary conditions to leading order, and in
this order it is equal to $F_c$, which in the present
parametrization reads
\begin{equation}\label{eq:ffc}
\F_{c}=\frac{L^2}{4\beta\PP(1-y)}\ ;
\end{equation}
as before, $(1-y)$ is the relative continuum power.

Like $F_p$, $\F_p$ is the negative of the maximal value of the exponent in Eq.\
(\ref{eq:zif}), subject to a given total power $y\PP$ and the
boundary conditions. However, the fixed boundary conditions break
the translation invariance, and the maximization is achieved when
pulses are created near the boundaries. We use standard
variational methods to obtain the result: $\F_p$ is the minimum
of four candidate function distinguished by the configuration of
the boundary pulses labelled by the four possible sign choices in
\begin{equation}\textstyle\label{eq:ffp}
\F_{p}=\frac{2\sqrt\beta}{3\alpha}\Big(2\lambda^{\frac32}
\pm\big(\lambda-\frac\alpha2|\psii|^2\big)^{\frac32}
\mp\big(\lambda-\frac\alpha2|\psif|^2\big)^{\frac32}\Big)
-\lambda y\PP
\end{equation}
In the first ambiguous sign the upper (lower) choice refers to the
possibility that a pulse maximum lies at $z>0$ ($z<0$). The
second sign choice refers similarly to the position of a pulse
maximum relative to the $z=L$ boundary. $\lambda$ is a Lagrange
multiplier for the power constraint, given implicitly by
\begin{equation}\textstyle
y\PP=\frac{2\sqrt\beta}\alpha\Big(2\sqrt\lambda\pm\sqrt{\lambda-\case12\alpha|\psii|^2}
\mp\sqrt{\lambda-\case12\alpha|\psif|^2}\Big)\ .
\label{eq:plambda}\end{equation} The total free energy $\F$ is
obtained by minimizing the sum $\F_\mathrm{p}+\F_\mathrm{c}$ with
respect to $\lambda$ and the possible choices of sign, i.e.,
$\F=\min\limits_\mathrm{\lambda,\pm,\pm}\F_c+\F_{p}$.

It is left to show that this expression for $\F$ satisfies the
differential equation (\ref{eq:fdiff}). We demonstrate that each
of the four possible sign choices satisfies Eq.\ (\ref{eq:fdiff}),
from which the result follows immediately. To this end we define
$\bar\lambda=\frac{L^2}{4\beta\PP\big(1-y(\bar\lambda)\big)^2}$, which is the value of $\lambda$ which minimizes $\F$ for a given choice of sign, as determined from Eqs.\ (\ref{eq:ffp}-\ref{eq:plambda})
The derivatives of $\F$ then read
\begin{equation}
\partial_\PP\F=
\bar\lambda\ , \qquad
\partial_{\psif}\F=\mp\psif^*\sqrt{\beta(\bar\lambda-\case12\alpha|\psif|^2)}\,,
\end{equation}
where the upper or lower sign choice corresponds to the second
ambiguous sign in (\ref{eq:ffp}-\ref{eq:plambda}). It is now
straightforward to verify that $\F$ and its derivatives satisfy
(\ref{eq:fdiff}) for all sign choices, as was claimed, finishing
the demonstration.


\begin{thebibliography}{10}
\bibitem{haus-review} H. A. Haus, ``Mode-Locking of Lasers",
IEEE J. Sel. Top. Quant. {\bf 6}, 1173 (2000)
\bibitem{GFPRL} A. Gordon and B. Fischer,
Phys. Rev. Lett. {\bf89}, 103901, (2002)
\bibitem{GGF03} O. Gat, A. Gordon, and B. Fischer, ``Solution of a statistical mechanics
model of passive mode locking'', Phys. Rev. E, in press, ArXiV preprint
\texttt{cond-mat/03011241}; (http://arxiv.org/cond-mat/0311241 ).
\bibitem{GFOC}  A. Gordon and B Fischer, Opt. Commun {\bf 223}, 151 (2003)
\bibitem{hh} P.C. Hohenberg and B.I. Halperin, Rev. Mod. Phys. \textbf{48}, 435 (1977)

\bibitem{stratonovich} R. L. Stratonovich, 
 in \emph{``Noise in nonlinear
dynamical systems''}, Vol. 1, edited by F. Moss and P. V. E.
McClintock, Cambridge University Press (1989)
\bibitem{experiment} A. Gordon, B. Vodonos, V. Smulakovski and B.
Fischer, Opt. Express {\bf 11}, 3418 (2003)

\end{thebibliography}
\end{document}